\documentclass[prd,twocolumn,nofootinbib,reprint]{revtex4-1}
\usepackage{amsmath,amssymb,amsthm,bm,braket,ascmac,bbm}
\usepackage{graphicx}
\usepackage{color}

\theoremstyle{definition}

\newcommand{\pc}{\ \mathrm{pc}}

\newcommand{\vecs}[1]{\mbox{\boldmath${#1}$}}

\begin{document}
\title{Axion production from primordial magnetic fields}
\author{Kohei Kamada$^1$ and Yuichiro Nakai$^{2}$}
\affiliation{\vspace{2mm} $^1$ School of Earth and Space Exploration, Arizona State University, Tempe, Arizona 85287, USA \\
$^2$Department of Physics, Harvard University, Cambridge, Massachusetts 02138, USA}

\begin{abstract}
\vspace{3mm}

Production of axionlike particles (ALPs) by primordial magnetic fields may have significant impacts on cosmology.
We discuss the production of ALPs in the presence of the primordial magnetic fields.  
We find a region of the ALP mass and photon coupling which realizes the observed properties of the dark matter
with appropriate initial conditions for the magnetic fields.
This region may be interesting in light of recent indications for the 3.5 keV lines from galaxy clusters.
For a small axion mass, a region of previously allowed parameter spaces is excluded by
overproduction of ALPs as a hot/warm dark matter component.
Since the abundance of ALPs strongly depends on the initial conditions of primordial 
magnetic fields, our results provide implications for scenarios of magnetogenesis.

\end{abstract}
\pacs{***}
\maketitle

\section{Introduction}\label{sec:intro}

Physics of the dark matter (DM) beyond the standard paradigm of the weakly interacting massive particle (WIMP) has recently enhanced its presence.
Despite intensive searches, any indication of the existence of the WIMP has not been found so far.
Moreover, in light of the naturalness of the electroweak symmetry breaking,
null results of the LHC experiments may suggest nonstandard signatures of supersymmetry (SUSY) such as R-parity violation
\cite{Barbier:2004ez,Graham:2014vya,Heidenreich:2014jpa} or stealth SUSY
\cite{Fan:2011yu,Fan:2012jf,Nakai:2015swg,Fan:2015mxp}
where the lightest supersymmetric particle cannot be the DM. 
If this is the case, another DM candidate is required.

Axionlike particles (ALPs) with very weak interactions and a tiny mass are promising alternatives to the WIMP idea
(for reviews, see \cite{Sikivie:2006ni,Marsh:2015xka}).
They may appear as pseudo-Nambu-Goldstone bosons (PNGBs) of some spontaneously broken global symmetries or by-products
of string theory compactifications.
Various production mechanisms of ALPs have been studied so far.
Thermal production of ALPs is likely to give a large free-streaming length and prevent structure formation~{\cite{Kolb:206230,Marsh:2015xka}}.
On the other hand, nonthermal production via the misalignment mechanism in the early Universe,
as discussed in \cite{Preskill:1982cy,Abbott:1982af,Dine:1982ah} for QCD axions,
\cite{Arvanitaki:2009fg,Acharya:2010zx,Marsh:2011gr,Higaki:2011me} for string axions
and \cite{Arias:2012az} for more general setups,
can give rise to the observed cold dark matter (CDM).
However, one possible problem in this production mechanism is a tight constraint from DM isocurvature perturbations
(see {\it e.g.} Ref.~\cite{Hamann:2009yf}).
This constraint gives an upper bound on the inflationary scale and excludes 
high-scale inflation models which can be tested in near-future observations.\footnote{
For possible solutions to this problem, see {\it e.g.} Refs.~\cite{Linde:1990yj,Linde:1991km,Dine:2004cq,
Kawasaki:2015lpf,Nomura:2015xil}.
}

Recently,  Fermi has observed a deficit of secondary GeV gamma rays
from TeV blazars
\cite{Neronov:1900zz,Tavecchio:2010mk,Ando:2010rb,Dolag:2010ni,Taylor:2011bn,Takahashi:2013lba,Chen:2014rsa,Finke:2015ona}. 
This observation can be explained by intergalactic magnetic fields (IGMFs) that broaden the secondary cascade photons, with a characteristic field strength $B_0 \gtrsim 10^{-19}$~G at Mpc scales (smaller scale IGMFs need a stronger strength to explain the deficit). 
Such IGMFs, if any, may have a primordial origin [primordial magnetic fields (PMFs)].\footnote{There are several 
proposals of magnetogenesis in the early Universe such as inflationary magnetogenesis~\cite{Turner:1987bw, Ratra:1991bn,Garretson:1992vt} (see also \cite{Fujita:2015iga,Adshead:2016iae})
and magnetogenesis from a strong first order phase transition~\cite{Hogan:1983zz,Quashnock:1988vs,Vachaspati:1991nm}. }
That is, we could imagine a scenario that
there exist strong magnetic fields in the very early Universe.
A natural question is then what the existence of the PMFs implies for the ALP DM paradigm.

In this article, we consider the production of ALPs via photon-axion conversion in the presence of PMFs. 
Photon-axion conversion is a process that has been well studied theoretically
\cite{Raffelt:1987im,Sikivie:1983ip} and discussed in different contexts
\cite{Ahonen:1995ky,Csaki:2001yk,Deffayet:2001pc,Long:2015cza}.
We show that a  sufficiently large number of ALPs could be produced in the early Universe, with a relatively long free-streaming length, via this conversion process.
We find a viable region of the ALP mass and photon coupling which predicts the appropriate properties for the DM such as its abundance and free-streaming length 
with suitable initial conditions for the PMFs. 
Moreover, for a small axion mass, a region of the previously allowed parameter space can be excluded by
the upper limit on the hot/warm component of the DM if the strength
of PMFs is relatively large but consistent with the present constraints on IGMFs. 

\section{Evolution of magnetic fields}\label{sec:scaling}

Before discussing the ALP production from PMFs, we first summarize the setup 
and assumptions on the cosmological evolution of PMFs.\footnote{Magnetic fields are generated in the early Universe as hypermagnetic fields and 
turn into (electro)magnetic fields at the electroweak phase transition/crossover. Assuming that 
the transition proceeds smoothly without a substantial change of the field strength, 
we do not distinguish the hyper gauge field from the electromagnetic field throughout the paper.} 
We here consider the case where nonhelical\footnote{In the maximally helical case, we would suffer from baryon overproduction~\cite{Fujita:2016igl,Kamada:2016eeb,Kamada:2016cnb}. For simplicity, we also do not consider the partially helical case.} PMFs are produced in the radiation dominated era
at a temperature $T_{\rm i}$ (or a time $t=t_{\rm i}$) 
with a causal process such as 
a strong first order phase transition in a hidden sector. 
The evolution of PMFs is described by magnetohydrodynamics (MHD) equations 
and is hard to evaluate in principle. 
However, it has been found that PMFs evolve according to a scaling law~\cite{Banerjee:2004df,Durrer:2013pga} 
until recombination
and after that evolve adiabatically. 
Here we assume that PMFs have a spectrum (blue at large scales) described by the 
characteristic field strength $B_p$ at the peak scale $\lambda_B$ which is 
identified as the correlation length and that they enter the scaling regime quickly 
after their generation. 
In the absence of late time entropy production, the strength of PMFs at a conformal time $\tau$ before recombination can be roughly written in terms of 
that of the present IGMFs,\footnote{To be precise, the scaling relation with a constant
$n_B$ applies only to the radiation dominated era with turbulent plasma and  
does not hold in all of cosmic history due to the neutrino or photon streaming effect
or matter domination. However, it turns out that the relation between the IGMFs and PMFs 
can be roughly evaluated as if the scaling relation holds until recombination. See the discussion in Refs.~\cite{Banerjee:2004df,Durrer:2013pga,Jedamzik:2010cy}.}
\begin{equation}
\begin{split}
	&B_p (\tau)= \left( \frac{a(\tau)}{a_0} \right)^{-2} \left( \frac{\tau}{\tau_{\rm rec}} \right)^{-n_B} B_0,
\label{eq:Blam_from_B0lam0}
\end{split}
\end{equation}
where $a(\tau)$ and $a_0$ are the scale factor at $\tau$ and today, respectively, 
and $\tau_{\rm rec}$ is the conformal time at recombination. 
$B_0$ is the characteristic strength of the present IGMFs.
In terms of the initial magnetic field strength $|\vecs{B}|=B_{\rm i}$, it is also written as
\begin{equation}
B_p(T)=B_{\rm i}(T/T_{\rm i})^{2+n_B} \label{Bevolution}
\end{equation} 
in the radiation dominated era.

The exponent $n_B$ 
in the scaling relations is subject to a controversy 
and differs by MHD simulations and analytical estimations. 
For example, in the direct cascade process, the exponent of the scaling law 
is obtained analytically~\cite{Banerjee:2004df,Durrer:2013pga,Jedamzik:2010cy} and also numerically~\cite{Banerjee:2004df}
as
\begin{equation}
{\rm (i)} \quad n_B=n/(2+n), 
\end{equation}
where $n$ is determined by the spectrum index of magnetic fields and fluid velocity fields.\footnote{$n \geq 3$ is required from the causality~\cite{Durrer:2013pga,Jedamzik:2010cy}.}
On the other hand, it has been recently claimed that an ``inverse transfer'' process would occur~\cite{Brandenburg:2014mwa,Zrake:2014mta} and the exponent of the scaling relation is identified as 
\begin{equation}
{\rm (ii)} \quad n_B=1/2.
\end{equation}
To be fair, we have both possibilities 
for the evolution of nonhelical magnetic fields in mind. 
Note that the resultant ALP abundance strongly depends on the 
strength of magnetic fields at their generation but is not very sensitive to 
the exponents of the scaling laws of the evolution, as we will see.  

In the above discussion, we have implicitly assumed that the correlation length 
is comparable to the largest processed eddy scale, $\lambda_B \sim v_A t$, 
with $v_A$ being the Alfv\'en velocity that depends on the magnetic field strength.
{This feature is observed in the MHD simulation~\cite{Banerjee:2004df,Kahniashvili:2012uj}}. 
Thus the comoving correlation length is fixed at recombination, which gives the linear relation
between the present strength and correlation length of the IGMFs as $\lambda_0 \simeq 1 \pc \times \left(B_0/10^{-14}\, {\rm G}\right)$~\cite{Banerjee:2004df}.\footnote{Although 
one must be careful for the treatment of Alfv\'en velocity at recombination that is in the matter dominated era, it turns out that the evaluation with the ``radiation domination approximation'' gives 
roughly correct relations~\cite{Banerjee:2004df,Durrer:2013pga,Jedamzik:2010cy}.}
In other words, if the MFs are causally generated in the early Universe, the correct property of the IGMFs is given on the line 
$\lambda_0 \simeq 1 \pc \times \left(B_0/10^{-14}\, {\rm G}\right)$ in the $B_0$-$\lambda_0$ plane.
The latest analysis of the TeV blazars by Fermi~\cite{Finke:2015ona} gives a constraint 
to explain the deficit of the GeV cascade photons. 
Since the $\lambda_0 \simeq 1 \pc \times \left(B_0/10^{-14}\, {\rm G}\right)$ line is in the ``ruled-out'' region of the $B_0$-$\lambda_0$ plane in Ref.~\cite{Finke:2015ona} for $B_0\lesssim 10^{-16}$~G and $\lambda_0\lesssim 10^{-2}$~pc, 
the constraint reads $B_0\gtrsim 10^{-16}$~G and $\lambda_0\gtrsim 10^{-2}$~pc 
if the IGMFs are generated in the early Universe by a causal mechanism so that they satisfy the linear relation between $B_0$ and $\lambda_0$. On the other hand, the upper bound on the IGMF strength at Mpc scales is given by 
the cosmic microwave background (CMB) as $B_0<10^{-9}$~G~\cite{Ade:2015cva}. 

\begin{figure}[!t]
  \begin{center}
   \includegraphics[clip, width=6.5cm]{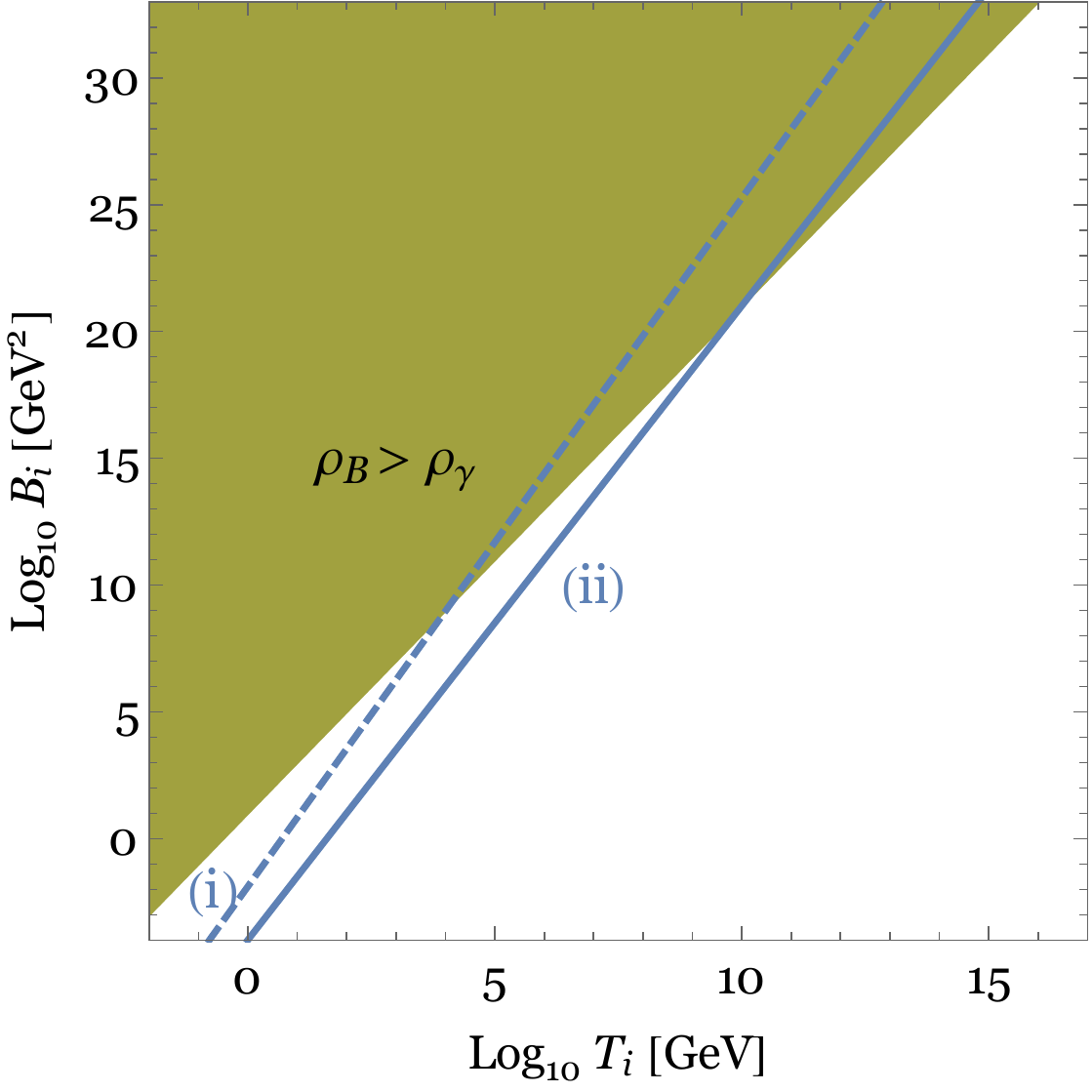}
  \end{center}
\vspace{-0.3cm}
  \caption{Constraints on the initial strength of PMFs $B_{\rm i}$ and temperature $T_{\rm i}$.
The (green) shaded region is excluded by $\rho_B(t_{\rm i})> \rho_\gamma(t_{\rm i})$.
The (blue) dashed and solid lines represent the initial conditions 
that predict $B_0=10^{-16}$~G for case (i) with $n=5$~\cite{Jedamzik:2010cy,Wagstaff:2014fla} and case (ii)
respectively. In the regions below the lines, 
the deficit of the GeV cascade photons from TeV blazars cannot be explained by the PMFs.  
In that region, the PMFs are free from CMB constraints. Note that the analytic investigation discussed in the text as well as the MHD simulation suggests that the present IGMF strength and correlation length must satisfy the linear relation, $\lambda_0 \simeq 1 \pc \times \left(B_0/10^{-14}\, {\rm G}\right)$~\cite{Banerjee:2004df}. By using this relation, we reinterpret the condition in Ref.~\cite{Finke:2015ona} in which the IGMFs can explain the blazar observation as $B_0\gtrsim 10^{-16}$~G and $\lambda_0 \gtrsim 10^{-2}$~pc.
  \label{fig:mag}}
\end{figure}

Since the energy density of PMFs decreases faster than that of radiation, 
the ratio between 
the energy density of PMFs $\rho_B=B_p^2/2$ 
and that of radiation is larger for higher temperatures. 
This, in turn, gives an upper bound on the initial strength of PMFs 
by requiring $\rho_B(t_{\rm i})<\rho_\gamma(t_{\rm i}) \sim T_{\rm i}^4$, 
depending on the scaling laws of the magnetic field evolution. 
Since we here do not specify the magnetogenesis mechanism,
this energy consideration uniquely  gives the upper bound on the initial field strength. 
Figure~\ref{fig:mag} shows constraints on the strength of magnetic fields $B_{\rm i}$ at the initial temperature $T_{\rm i}$ for
each case of the exponents of the scaling laws.
Note that the initial strength of PMFs can be expressed in terms of the 
present IGMF strength and the initial temperature through Eq.~\eqref{eq:Blam_from_B0lam0}. 
We can see that 
upper bounds on the initial strength of PMFs and temperature 
that can explain the deficit of the GeV cascade photons from blazars are given 
as $B_{\rm i}^{1/2}<  10^4 \, (10^{10}) \, {\rm GeV}$ and $T_{\rm i}< 10^4 \, (10^{10}) \, {\rm GeV}$ for case (i) with $n=5$~\cite{Jedamzik:2010cy,Wagstaff:2014fla} (for case (ii)). 
Hereafter we use these values as references.

\section{Axion production}\label{sec:production}

We here discuss how ALPs are produced by PMFs.
ALPs are PNGBs of some global symmetries and couple to matter and gauge bosons only derivatively.
The coupling of an ALP $\phi$ to the electromagnetic field is given by
\begin{equation}
\mathcal{L}_{\phi} \supset -\frac{1}{4} g \phi F_{\mu\nu}{\tilde F}^{\mu\nu}=g \phi \vecs{E} \cdot \vecs{B}, \quad
g \simeq \frac{\alpha}{2 \pi} \frac{1}{f_\phi}, \label{EB}
\end{equation}
where $F_{\mu\nu}$ is the electromagnetic field strength tensor, 
${\tilde F}^{\mu\nu} \equiv \frac{1}{2}\epsilon^{\mu\nu\rho\sigma}F_{\rho \sigma}$ is its dual, 
and $\vecs{E}$ and  $\vecs{B}$  are the electric and magnetic fields, respectively.
$\alpha \equiv e^2 / 4\pi$ is the fine structure constant
and $f_\phi$ is the decay constant. 
We can see that in the presence of a background magnetic field this coupling induces a mixing between an ALP and the electric field whose
polarization is parallel to the magnetic field.

The probability of photon-axion conversion, $\gamma \rightarrow \phi$,
in the presence of plasma has been studied in Ref.~\cite{Deffayet:2001pc}
where the linearized field equations of the photon-axion system with an energy $\omega$ are given by
\begin{equation}
\begin{split}
(\omega - i \partial_y + \mathcal{M})
\begin{pmatrix}
| \gamma \rangle \\
| \phi \rangle
\end{pmatrix} = 0,
\quad \mathcal{M} =
\begin{pmatrix}
\Delta_\gamma & \Delta_M \\
\Delta_M & \Delta_\phi
\end{pmatrix}. \label{fieldeq}
\end{split}
\end{equation}
Here, the coordinate $y$ is along the direction of propagation and we denote the state vectors of the photon
and the ALP as $| \gamma \rangle$, $| \phi \rangle$.
For the matrix $\mathcal{M}$, $\Delta_\phi \equiv - m_\phi^2 / 2 \omega$ comes from the ALP mass $m_\phi$
and $\Delta_M \equiv g \langle {\bm e} \cdot {\bm B} \rangle / 2 \simeq g B_p / 2$ originates from the photon-axion mixing in the presence of the magnetic field. Here, ${\bm e}$ denotes the photon polarization vector and the bracket represents the ensemble average.
The upper left component comes from the nonzero effective photon mass in magnetized plasma. 
In the present case, the effective mass is dominated by the Debye mass,
$m_D^2 \sim e^2 T^2$, and we obtain $\Delta_\gamma \simeq - m_D^2/2 \omega$.
From the field equations \eqref{fieldeq}, the photon-axion conversion probability after a distance of propagation $y =L$ is given by~\cite{Deffayet:2001pc}
\begin{equation}
\begin{split}
&P (\gamma \rightarrow \phi) = (\Delta_M  L)^2 \, \frac{\sin^2 (\Delta_{\rm osc} L /2) }{ (\Delta_{\rm osc} L /2)^2}, \\[1.5ex]
&\qquad \,\, \Delta_{\rm osc}^2 \equiv (\Delta_\gamma - \Delta_\phi )^2 + 4 \Delta_M^2 .
\end{split}
\end{equation}
For the parameter range of interest, $T\gtrsim B \gg m_\phi$, we approximately find $\Delta_{\rm osc} \simeq \Delta_\gamma$.

We now derive the kinetic equation for ALPs from the photon-axion conversion probability 
obtained above.
Here we can use the same procedure in the case of production of sterile neutrinos via oscillations
\cite{Barbieri:1989ti,Barbieri:1990vx,Kainulainen:1990ds,Enqvist:1990ad,Enqvist:1991qj,Dodelson:1993je,Stodolsky:1986dx,Foot:1996qc,Abazajian:2001nj}.
Following the discussion of~\cite{Foot:1996qc,Abazajian:2001nj} (see also Refs.~\cite{Barbieri:1989ti,Barbieri:1990vx,Kainulainen:1990ds,Enqvist:1990ad,Enqvist:1991qj,Dodelson:1993je}),
the photon-axion conversion rate in a unit time is evaluated in terms of the probability averaged over photons in the ensemble,
\begin{equation}
\Gamma (\gamma \rightarrow \phi) = \frac{\Gamma_\gamma}{2} \langle P (\gamma \rightarrow \phi) \rangle,
\end{equation}
where $\Gamma_\gamma \sim \alpha^2 T$ is the thermally averaged collision rate of photons.
This expression of the conversion rate can be understood as follows.
The collision of a photon leads to collapse of the photon wave function into either a pure photon eigenstate or a pure ALP eigenstate.
Then, the collision is a measurement.
The rate of the measurements is given by $\Gamma_\gamma$
and the origin of the factor $1/2$ has been discussed in Ref.~\cite{Stodolsky:1986dx}.
In the ensemble average of the probability, we take $\langle \omega \rangle \sim T$ and 
$\langle \sin^2 (\Delta_{\rm osc} L /2) \rangle \rightarrow 1/2$ since 
$\Delta_{\rm osc} L \simeq \Delta _{\rm osc} \Gamma_\gamma^{-1} \gg 1$.

In the radiation dominated Universe, the evolution of 
the number-to-entropy ratio of ALPs, $\eta_\phi \equiv n_\phi/s$
($n_\phi$ is the number density of ALPs and $s=2 \pi^2 g_{*s}T^3/45$ is the entropy density), obeys the following 
kinetic equation:
\begin{equation}
\frac{d \eta_\phi}{dt}=\Gamma (\gamma \rightarrow \phi) \frac{n_\gamma-n_\phi}{s} = c g^2 \frac{ B_p^2}{T} \left(1-\frac{n_\phi}{n_\gamma}\right), \label{kinetic}
\end{equation}
where $n_\gamma$ is the number density of photons and 
we have introduced a numerical factor $c = \mathcal{O} (0.1) $.

If $H \gg \Gamma(\gamma \rightarrow \phi)$ is satisfied throughout the cosmic history, we obtain the ALP number-to-entropy ratio by integrating the kinetic equation with neglecting $n_\phi/n_\gamma$ as
\begin{equation}
\begin{split}
\eta_\phi(t)&=\int^t_{t_{\rm i}} dt' \, cg^2  \frac{B_{\rm i}^2(T(t')/T_{\rm i})^{2(2+n_B)}}{ T(t')} 
\simeq c' g^2 \frac{B_{\rm i}^2 M_{\rm pl}}{  T_{\rm i}^3},
\end{split}
\end{equation}
where $c'$ is a numerical factor with 
$\mathcal{O}(0.01)$. Here we have used Eq.~\eqref{Bevolution} and 
$t= (2H)^{-1} \propto T^{-2}$ in the radiation dominated Universe. 
Since the photon-axion conversion rate decreases quickly, $\eta_\phi$ is fixed just after the magnetic field generation. 
The present ALP energy-to-entropy ratio is then given by 
\begin{equation}
\begin{split}
\frac{\rho_{\phi, B}}{s}&\simeq \, 2 \times  10^{-{10}} \, {\rm GeV} \times {\left(\frac{c'}{0.01}\right)} \left(\frac{m_\phi}{{10} \, {\rm keV}}\right)
 \\
\quad\times &\left(\frac{B_{\rm i}}{(10^{11} \, {\rm GeV})^2}\right)^2\left(\frac{T_{\rm i}}{10^{11} \, {\rm GeV}}\right)^{-3} \left(\frac{g}{10^{-16} \, {\rm GeV^{-1}}}\right)^{2}.  \label{APCabundance}
\end{split}
\end{equation}

If once $H \ll \Gamma(\gamma \rightarrow \phi)$ is satisfied in the cosmic history, 
ALPs are thermalized through the photon-axion conversion process
and the present ALP energy-to-entropy ratio is evaluated as 
\begin{equation}
\frac{\rho_{\phi,B}}{s} = \frac{m_\phi n_\phi }{s} \simeq  2.6 \times 10^{-9} \, {\rm GeV} \times \left(\frac{m_\phi}{1\, {\rm keV}}\right). 
\end{equation}
Note that the thermalized axion number density to entropy ratio is fixed as $n_\phi/s=(\zeta(3)/\pi^2)T^3/((2\pi^2/45)g_{*s} T^3) \simeq 2.6 \times 10^{-3}$, 
where we take $g_{*s}$ to be the value for all the Standard Model particles, $g_{*s} = 106.75$.

Since the present DM abundance in the Universe is given by $\rho_{\rm DM}/s {=\Omega_{\rm DM} \rho_c /s}\simeq 4.0 \times 10^{-10} \, \rm GeV$~{\cite{Ade:2015xua,Kamada:2011ec}}, the fiducial values of parameters in Eq.~\eqref{APCabundance} can explain the abundance of the DM. Note that it is difficult to explain both the ALP DM abundance and 
the deficit of the GeV cascade photons from TeV blazars simultaneously, see Fig.~\ref{fig:mag}.
One possible solution is to consider the case where the PMFs evolve adiabatically first 
and start the direct cascade or the inverse transfer at a later time similar to the situation 
discussed in Ref.~\cite{Fujita:2016igl}. 
For example, consider the case where  
the magnetic fields are produced at $T=10^{11}$~GeV with 
$B_{\rm i}=10^{22} \ {\rm GeV}^2$ and evolve adiabatically ($n_B=0$) for a while. 
The ALP abundance fixed at the $T=10^{11}$~GeV and it coincides with the present DM abundance. 
If the eddy scale of turbulent plasma gets comparable to the coherent length of the magnetic fields 
at $T\simeq 10^9$~GeV with $B_p \simeq 10^{18}~{\rm GeV}^2$ and they evolve with the 
scaling law [(ii); $n_B=1/2$] after that, the present strength and coherent length of IGMFs 
are $B_0 \sim 10^{-16}$~G and $\lambda_B \sim 0.01$~pc. 
Thus, the axion DM abundance and the blazar observation 
can be explained simultaneously.
However, this scenario needs a new parameter (the initial correlation length 
or the time when the scaling law changes) and hence we have less predictive power. 
Moreover, unfortunately, we do not know any concrete magnetogenesis mechanisms 
that lead to this scenario.

For ALPs produced by primordial magnetic fields to explain the present Universe,
we need to know if the produced ALPs are stable and cold enough.
ALPs can decay into two photons, $\phi \rightarrow \gamma \gamma$, through the photon coupling~\eqref{EB}.
The lifetime is given by
\cite{Arias:2012az}
\begin{equation}
\begin{split}
\tau_\phi \simeq 10^{28} \, {\rm s}
\times \left( \frac{g}{10^{-16} \, \rm GeV^{-1}} \right)^{-2} \left( \frac{m_\phi}{1 \, \rm keV} \right)^{-3} . \label{lifetime}
\end{split}
\end{equation}
In the parameter range where ALPs produced by photon-axion conversion can explain the observed DM abundance,
the ALP lifetime is easily much longer than the age of the Universe, $t \sim 10^{17} \, \rm s$.

The ``temperature'' of ALPs produced by this mechanism, $T_\phi$, is the same to the photon at production and is just redshifted afterwards. 
The comoving free-streaming horizon at matter radiation equality 
is estimated as~\cite{Boyarsky:2008xj}
\begin{equation}
\begin{split}
\lambda_{\rm FS} \simeq &\,\,1 \, {\rm Mpc} \times \left( \frac{m_\phi}{1 \, \rm keV} \right)^{-1}\left( \frac{T_\phi/T} {0.33}\right) \\
&\times \left( \, 1+0.03 \log \left[ \left(\frac{m_\phi}{1 {\rm keV}}\right) \left(\frac{0.1}{\Omega_{\rm DM} h^2}\right) \left(\frac{0.33}{T_\phi/T} \right)\right] \,  \right) ,
\end{split}
\end{equation}
where  
$\Omega_{\rm DM} h^2$ is the present density parameter of the DM.\footnote{The estimate is the same as that in the case where ALPs are once thermalized.} 
We here take into account the change in the effective numbers of relativistic degrees of freedom of the Standard Model sector, $T_\phi /T = \left( \frac{3.91}{106.75} \right)^{1/3} \simeq 0.33$.  
A large free-streaming length prevents structure formation and is constrained by Lyman-$\alpha$ forest observations
\cite{Boyarsky:2008xj,Harada:2014lma,Kamada:2016vsc}.
The upper bound of the length is around 1 Mpc and hence
ALPs with $\lambda_{\rm FS} > 1 \, \rm Mpc$ cannot be the main component of the DM.
For $m_\phi \lesssim 1 \, \rm keV$, the constraint on the ratio of the energy density of ALPs in the total DM density $\rho_{\rm DM}$
is given by $\rho_{\rm ALP} / \rho_{\rm DM} < 0.6$ \cite{Boyarsky:2008xj}.
For $m_\phi \ll 1 \, \rm keV$, the Planck 2015 temperature and polarization data give a stronger bound, $\rho_{\rm ALP} / \rho_{\rm DM} < 0.3 \times 10^{-2}$, 
interpreting the constraint of Ref.~\cite{DiValentino:2015wba} in terms of the ALP energy density. 
We will see that these constraints exclude a region of the parameter space of ALPs.


Here we comment on other ALP production mechanisms.
If the ALP $\phi$ stays at a different field point from the potential minimum during inflation,
it begins to oscillate around the minimum of the potential when the Hubble parameter 
becomes comparable to the ALP mass.
The oscillation behaves as a matter which survives until today, which is called 
the misalignment mechanism. 
The present energy-to-entropy ratio of ALPs produced by this mechanism is given by~\cite{Arias:2012az} 
\begin{equation}
\begin{split}
\frac{\rho_{\phi, \phi_1}}{s} \sim 10 \, {\rm GeV} \times \left( \frac{m_\phi}{1 \, \rm keV} \right)^{1/2}
\left( \frac{\phi_1}{10^{16} \, \rm GeV} \right)^2,
\end{split}
\end{equation} 
where $\phi_1$ is the initial value of the ALP field and we have assumed that 
the ALPs start oscillation in the radiation dominated era and the ALP mass at the time when the oscillation starts
is the same as the present mass $m_{\phi}$.
Natural values of $\phi_1$ are around $f_\phi$ and, 
for a large decay constant, we often suffer from the ALP overproduction problem. 
The correct DM abundance (or much less ALP abundance) can be also obtained, however,  by tuning the initial 
condition.
The degree of tuning is estimated by $\Delta_\phi \equiv \phi_1 / f_\phi$.
The produced ALPs can give the CDM with a small free-streaming length.
Note that they often generate
too large DM isocurvature perturbation in the case of high-scale inflation. 

ALPs are also produced via scattering of quarks and gluons in thermal equilibrium
such as $gg \rightarrow g \phi$.
If the temperature of the Universe is higher than the decoupling temperature of the scattering, 
\begin{equation}
T_D \sim 10^6 \, {\rm GeV} \times \left(\frac{g}{10^{-10} \, {\rm GeV^{-1}}}\right)^{-2}, 
\end{equation} 
ALPs are thermalized, which may cause an ALP overproduction problem.
Even if the temperature is below the decoupling temperature,
ALPs are still produced like gravitinos or freeze-in DM scenarios. 
The abundance of relic ALPs is evaluated as~\cite{Masso:2002np,Sikivie:2006ni,Graf:2010tv}
\begin{equation}
\begin{split}
\frac{\rho_{\phi, \rm th}}{s} \sim & \, 10^{-16} \, {\rm GeV} \times \left(\frac{m_\phi}{1 \, {\rm keV}} \right) \\
&\quad \times \left(\frac{g}{10^{-16} \, {\rm GeV}}\right)^{2} \left(\frac{T_{\rm R}}{10^{11} \, {\rm GeV}}\right),
\end{split}
\end{equation}
where $T_{\rm R}$ is the reheating temperature. 
We find that the number of thermally produced ALPs is smaller than that of ALPs produced 
from photon-axion conversion in the parameter range of interest.

\section{ALP dark matter}\label{sec:DM}

\begin{figure}[!t]
  \begin{center}
   \includegraphics[clip, width=6.5cm]{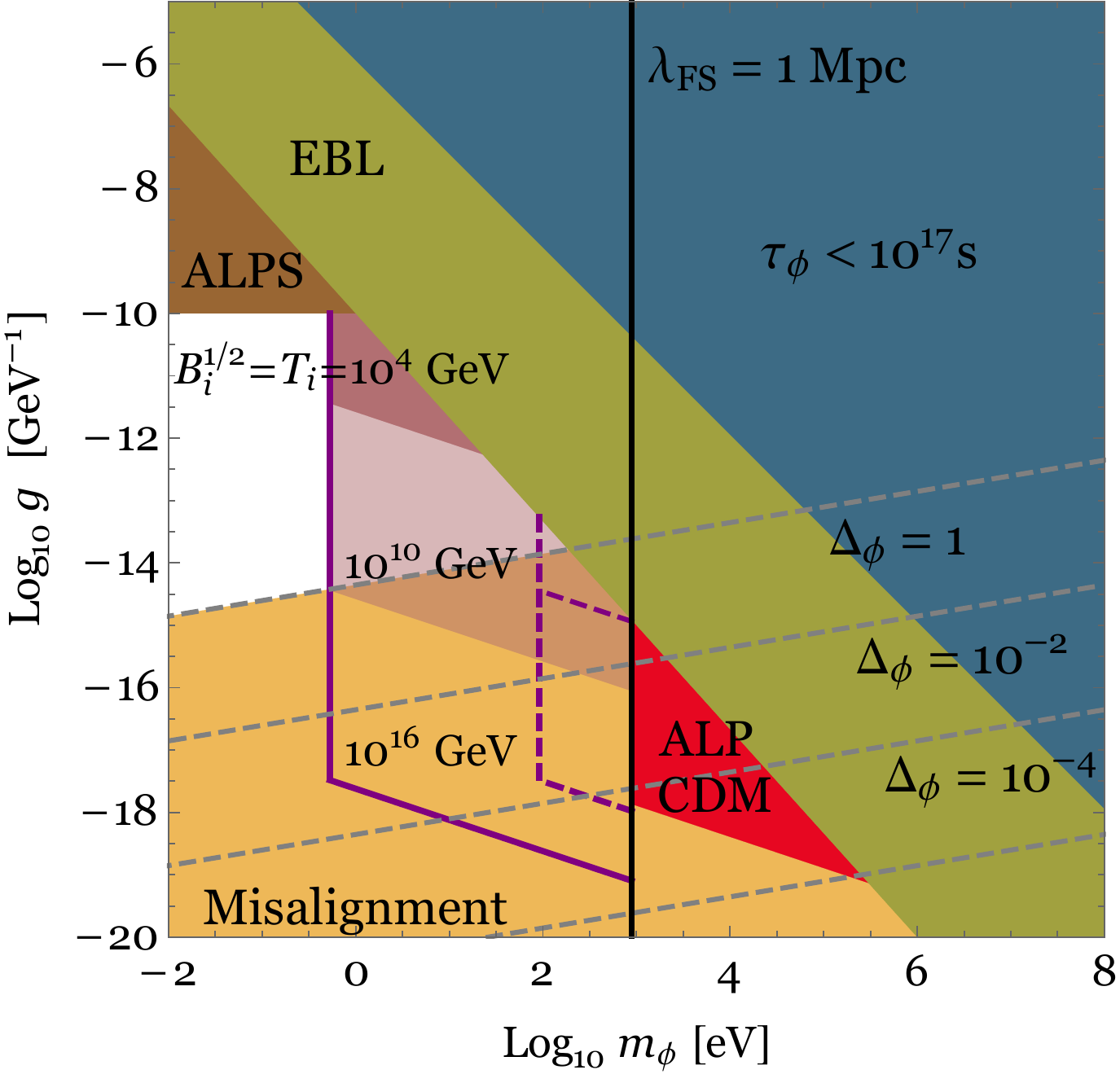}
  \end{center}
\vspace{-0.3cm}
  \caption{The constraints on the parameter space of the ALP mass $m_\phi$ and the photon coupling $g$ are shown. 
The blue region ($\tau_\phi<10^{17}$s),  green region  (EBL) and brown region (ALPS) are constrained by the ALP stability, astrophysical observations such as the EBL, and direct experimental and observational bounds on the photon coupling, respectively. 
The yellow region (Misalignment) can realize the correct abundance of the ALP DM produced by the misalignment mechanism
with an appropriate tuning of the initial amplitude $\phi_1$. The degree of tuning is shown by the gray dashed lines.
 The black solid line gives $\lambda_{\rm FS} = 1 \, \rm Mpc$.
For $m_\phi \gtrsim 1$ keV, ALPs can be the CDM. 
The red region (ALP CDM) gives the correct abundance of the ALP DM produced by primordial magnetic fields for $T_{\rm i}=B_{\rm i}^{1/2}<10^{16}$~GeV. 
For $m_\phi \lesssim 1$~keV, the parameter space can be constrained by the hot/warm DM, $\rho_{\rm ALP}/\rho_{\rm DM}<0.3 \times 10^{-2}$. 
The light (thick) purple region is ruled out for $T_{\rm i} = B_{\rm i}^{1/2}=10^{10} \, (10^{4}) \, \rm GeV$. 
The region above the solid purple line is ruled out for $T_{\rm i} = B_{\rm i}^{1/2}=10^{16} \, \rm GeV$.  
The regions surrounded by the dashed purple lines are ruled out by a conservative constraint $\rho_{\rm ALP}/\rho_{\rm DM}<0.6$.
}
  \label{fig:gmphi}
\end{figure}

We now investigate
the parameter space of the ALP mass and photon-ALP coupling 
and identify a region where ALPs produced by photon-axion conversion
can be responsible for the present DM.
We also give constraints from overproduction of the ALP hot/warm DM. 

Let us first summarize the known constraints on the ALP parameters. 
The first constraint for ALPs as the DM comes from its stability. 
As we have mentioned, the lifetime of ALPs $\tau_\phi$ [Eq.~\eqref{lifetime}] must be larger than the 
age of the Universe, $t\sim 10^{17}$ s.
Even if this is satisfied, 
partial decays of ALPs might cause phenomena inconsistent with astrophysical observations, such as the extragalactic background light (EBL) and 
extragalactic x-rays. 
These observations lead to a constraint on the ALP mass and the photon coupling, which is roughly given as
\cite{Arias:2012az}
\begin{equation}
\begin{split}
g < 10^{-10} \, {\rm GeV^{-1}} \times \left( \frac{m_\phi}{1 \, \rm eV}  \right)^{-5/3}.
\end{split}
\end{equation}
Furthermore, there are direct experimental and observational bounds on the photon coupling. 
The constraints come from
the light-shining-through-walls experiment ALPS and the helioscopes CAST and SUMICO.
Combining with the constraint from the short lifetime of ALPs, it is required to satisfy 
$
g < 10^{-10} \, {\rm GeV^{-1}}
$~\cite{Arias:2012az}.

Now we explore the possibility of the ALP dark matter. 
From Eq.~\eqref{APCabundance}, we can see if ALPs 
produced by photon-axion conversion through the PMFs can be the CDM for 
$m_\phi\gtrsim 1$ keV. 
Since the ALP abundance is proportional to $B_{\rm i}^2 $ and $T_{\rm i}^{-3}$ 
and energy condition gives a constraint $B_{\rm i}^{1/2} \lesssim T_{\rm i}$, 
the large ALP abundance is obtained when $B_{\rm i}^{1/2} \simeq T_{\rm i}$ 
with large temperature at production. 
The red region in Fig.~\ref{fig:gmphi}  represents 
the parameter space of the ALP mass $m_\phi$ and the photon coupling $g$
where ALPs can be the CDM. 
Here we take the highest value of 
the initial temperature and the square root of the magnetic field strength as the possible highest temperature of the Universe,
$T_{\rm i}=B_{\rm i}^{1/2}<10^{16} \, \rm  GeV$. 
The lowest initial temperature and the square root of the initial magnetic field strength 
that can explain the present DM are $T_{\rm i}=B_{\rm i}^{1/2}\simeq 10^{10-11}$ GeV.
Therefore, unfortunately, it is difficult to explain the CDM and the deficit of the GeV cascade photons from blazars simultaneously (See Fig.~\ref{fig:mag}).\footnote{
As we have mentioned in Sec.~\ref{sec:production}, if we consider the case where the PMFs evolve adiabatically first and start the direct cascade or the inverse transfer at a later time, 
one can construct a scenario where the blazar observation and the DM abundance are explained
simultaneously. However, this scenario needs another phenomenological parameter, which loses the 
one-to-one correspondence between the DM abundance and the present IGMF strength and 
hence weakens the predictive power.}
However, it should be emphasized that IGMFs can be explained by other mechanisms that 
occur at some later time 
and hence weaker PMFs are not worrisome. 
Note that, in the region where ALPs produced by PMFs can explain the CDM, those produced by the misalignment mechanism must be suppressed somehow strangely.  But it has a benefit that we do not suffer from too large DM isocurvature 
perturbation. Interestingly, it has been known that the collisionless CDM addresses some disagreements between the observations and numerical simulations of the galactic halos such as the ``core-cusp'' problem or ``too big to fail'' problem.
The ALP DM produced by photon-axion conversion via PMFs with free-streaming length of $\mathcal{O} (0.1$ - $1) \, \rm Mpc$ might be the 
candidate to resolve the issues. 

For $m_\phi \lesssim 1 \, \rm keV$, we can constrain the ALP parameter 
space from the hot/warm DM argument, $\rho_{\rm ALP}/\rho_{\rm DM}<0.3 \times 10^{-2}$. 
The light (thick) purple regions in Fig.~\ref{fig:gmphi} are excluded due to 
the overproduction of hot/warm ALPs for $T_{\rm i}=B_{\rm i}^{1/2}\simeq 10^{10} \, (10^{4}) \, \rm  GeV$.
These choices of the parameters $T_{\rm i}=10^{4}$ GeV and $10^{10}$ GeV predict IGMFs with $B_0=10^{-16}$ G today that can explain the deficit 
of the GeV cascade photons from blazars for the scaling laws in case (i) with $n=5$ 
and  case (ii), respectively (see Fig.~\ref{fig:mag}).
The region above the thick purple line is ruled out for $T_{\rm i}=B_{\rm i}^{1/2}\simeq 10^{16}$ GeV, the possible highest temperature of the Universe. 
We also show the parameter space excluded by the constraint $\rho_{\rm ALP}/\rho_{\rm DM}<0.6$, a more conservative one, with the dashed purple lines. 
Note that for $g>10^{-17}, 10^{-14}, 10^{-11} \, {\rm GeV}^{-1}$ ALPs are thermalized
for $T_{\rm i}=10^{16}, 10^{10}, 10^{4} \, \rm  GeV$ respectively and 
the region $m_\phi<0.5 \, (10^2) \, \rm  eV$ is not constrained by  $\rho_{\rm ALP}/\rho_{\rm DM}<0.3 \times 10^{-2} \, (0.6)$. 
Magnetogenesis models can be also constrained if future experiments 
identify ALPs within that parameter region.

\section{Discussions}\label{sec:conclusion}

We have discussed ALP production in the presence of PMFs 
without specifying their origin
and explored several features of the produced ALP. 
We have found a region of the ALP mass and photon coupling where ALPs produced via photon-axion conversion
provide the correct abundance of the DM.
In particular, the mass region for the ALP CDM via PMFs covers the point in Ref.~\cite{Jaeckel:2014qea}
that may explain the  recently indicated  emission lines at 3.5 keV from galaxy clusters and the Andromeda galaxy~\cite{Bulbul:2014sua,Boyarsky:2014jta}. 
These ALPs are different from those produced via the misalignment mechanism
in that the free-streaming length is relatively long.
If ALPs are detected in future observations, it might be possible to identify the dominant production mechanism
by this feature.

It should be emphasized that the origin and the evolution of PMFs are 
still under discussion. 
A strong first order phase transition~\cite{Hogan:1983zz,Quashnock:1988vs,Vachaspati:1991nm} can be a possible origin of PMFs, but at present we do not have a candidate 
for such phase transition. 
Inflationary magnetogenesis~\cite{Turner:1987bw, Ratra:1991bn,Garretson:1992vt}
is another option, but there are not satisfactory models (except for Ref.~\cite{Fujita:2016qab}).  
Although we have taken simple scaling laws of the evolution of PMFs for simplicity, 
their nature is not fully understood yet~\cite{Banerjee:2004df,Durrer:2013pga,Jedamzik:2010cy,Brandenburg:2014mwa,Zrake:2014mta}.   
Note that we did not consider the possibility of backreaction to PMFs from
photon-axion conversion. This might also change the evolution of PMFs
and make it possible for the PMFs to be the origin of the IGMFs responsible 
for the deficit of GeV cascade photons from blazars and at the same time provide the ALP DM.
The thorough treatment of this effect is left for a future investigation.

Our results give new implications for the nature of PMFs as well as ALPs.
As a future investigation, it is worth exploring to construct a magnetogenesis mechanism 
with $T_{\rm i} \simeq B_{\rm i}^{1/2} > 10^{11}$ GeV 
for the consistent ``ALP DM via PMF'' scenario. 
Once we determine the strength and correlation length of the IGMFs
by future gamma-ray observations, 
it will be interesting to construct a magnetogenesis model,
to perform careful studies on evolution of PMFs and to find if
the ALP DM and the IGMFs can be explained simultaneously.
This will also identify the ALP mass and photon-axion coupling strength. 
For a small ALP mass, we have also pointed out the possible problem that hot/warm relic ALPs can constrain
the properties of both PMFs and ALPs. 
If small mass ALPs will be detected, the strength of PMFs is constrained. 

\section*{Acknowledgements}

We would like to thank Paola Arias, 
Francis-Yan Cyr-Racine, Andrew Long and Tanmay Vachaspati 
for the useful discussions and comments.
K.K. acknowledges support from the U.S. DOE for this work under Grant No. DE-SC0013605
and would like to thank Harvard University for their kind hospitality where this work was initiated.
Y.N. is supported by a JSPS Fellowship for Research Abroad.

\appendix

\bibliography{ref}

\end{document}